\documentclass[aps, prx, superscriptaddress, reprint, twocolumn, floatfix, preprintnumbers,nofootinbib]{revtex4-2}
\usepackage{latexsym, amsmath, amsfonts, amssymb}

\usepackage{graphicx}

\usepackage{comment}

\usepackage{hyperref}
\hypersetup{colorlinks = true, linkcolor = blue, anchorcolor = blue, citecolor = blue, filecolor = blue, urlcolor = blue}
\usepackage{bbm}
\usepackage{mathrsfs}
\usepackage{array} 
\usepackage{xcolor}

\usepackage{tikz}
\usepackage{tikz-3dplot}

\newcommand{\be}{\begin{equation}} \newcommand{\ee}{\end{equation}}
\newcommand{\bea}{\begin{equation} \begin{aligned}} \newcommand{\eea}{\end{aligned} \end{equation}}

\newcommand{\cA}{\mathcal{A}}

\newcommand{\cC}{\mathcal{C}}

\newcommand{\cH}{\mathcal{H}}
\newcommand{\cI}{\mathcal{I}}

\newcommand{\cL}{\mathcal{L}}

\newcommand{\cN}{\mathcal{N}}
\newcommand{\cO}{\mathcal{O}}

\newcommand{\cR}{\mathcal{R}}

\newcommand{\cT}{\mathcal{T}}

\newcommand{\bR}{\mathbb{R}}

\newcommand{\bZ}{\mathbb{Z}}

\newcommand{\fu}{\mathfrak{u}}

\newcommand{\unit}{\mathbbm{1}}
\newcommand{\calD}{\mathscr{D}}

\begin{document}

\title{Defect Conformal Manifolds from
Phantom (Non-Invertible) Symmetries}

\newcommand{\OXFORD}{\affiliation{Mathematical Institute, University
of Oxford, Woodstock Road, Oxford, OX2 6GG, United Kingdom}}

\newcommand{\ULB}{\affiliation{Physique Theorique et Mathematique and International Solvay Institutes
Universite Libre de Bruxelles, C.P. 231, 1050 Brussels, Belgium}}

\newcommand{\IHES}{\affiliation{Institut des Hautes Etudes Scientifiques, 91440 Bures-sur-Yvette, France
}}

\author{Andrea Antinucci}
\OXFORD

\author{Christian Copetti}
\OXFORD

\author{Giovanni Galati}
\ULB

\author{Giovanni Rizi}
\IHES

\begin{abstract}
We explore a general mechanism that allows (1+1)d CFTs to have interesting interface conformal manifolds even in the absence of any continuous internal symmetry or supersymmetry. This is made possible by the breaking of an enhanced continuous symmetry--which is generically non-invertible-- arising in the folded theory. We provide several examples and showcase the power of the symmetry-based approach by computing the evolution of the reflection coefficient along the defect conformal manifold. We also discuss higher-dimensional generalizations and we comment on no-go theorems.
\end{abstract}

\maketitle

\section{Introduction}
Conformal manifolds are a rare and intriguing phenomenon. In CFTs with (spacetime) dimension greater than two, their occurrence is typically restricted to free or supersymmetric theories. Generically, naively marginal deformations acquire anomalous dimensions in conformal perturbation theory, becoming irrelevant or relevant. Only under exceptional mechanisms---often involving supersymmetry---a deformation can remain exactly marginal, preserving conformal invariance and giving rise to a conformal manifold.

The situation differs markedly for defects and boundaries of a CFT \cite{Padayasi:2021sik,Drukker:2022pxk,Herzog:2023dop}. In the presence of a $p$-dimensional defect $\calD$, a continuous global symmetry $G$ may be locally broken on the defect, modifying the Ward identity for the conserved Noether current $J^{\mu}$:
\be \label{eq: tilt}
\partial_\mu J^\mu = t(x) \, \delta_p(\Sigma_\calD) \, , \ \ \ t(x) = t_a(x) T^a \, , \ \  T^a \in \mathfrak{g} \, .
\ee
The operator appearing on the r.h.s of \eqref{eq: tilt} is called the \emph{tilt} operator. Conformal invariance imposes $\Delta_t = p$, and exponentiation of $t(x)$ generates a defect conformal manifold.\footnote{Further studies involving the tilt operator are \cite{Bartlett-Tisdall:2023ghh,Copetti:2023sya,Cuomo:2023qvp,Trepanier:2023tvb}.} Defect conformal manifolds are thus generic in theories with continuous symmetries, their existence being protected by the symmetry-breaking pattern due to the defect.
As two points on the conformal manifold are connected by a symmetry action, the underlying geometry is homogeneous \cite{Herzog:2023dop} and several interesting properties of the defect--such as the $g$-function \cite{Affleck:1991tk} and the reflection coefficient $\cR$ \cite{Quella:2006de}-- remain constant.

In this Letter, we explore a distinct scenario. Even when the bulk theory lacks continuous symmetries, the spectrum of co-dimension one conformal defects---or more generally, interfaces---can include nontrivial conformal manifolds. This appears to conflict with the earlier reasoning, as no evident mechanism protects the marginality of the defect operator. We argue, however, that such manifolds arise due to additional global symmetries appearing after folding the theory, which we term \emph{phantom} symmetries.

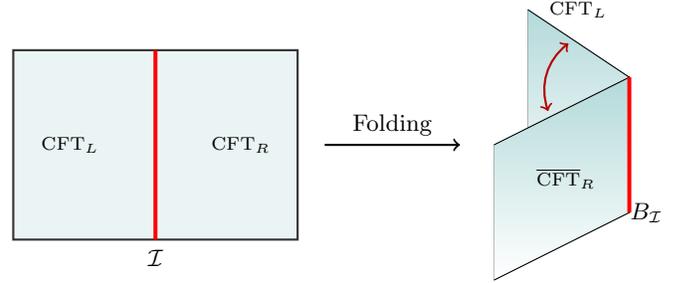
\begin{figure}
    \centering
    \begin{tikzpicture}[scale=0.9]

\begin{scope}[shift={(0,0)}, scale=0.7]

    \draw[thick,fill=white!90!teal, opacity=0.8] (-3,-2) rectangle (3,2);

    \node at (-1.8,0) {\scriptsize CFT$_L$};
    \node at (1.8,0) {\scriptsize CFT$_R$};
  
    \draw[red, line width=1.5] (0.,-2) node[below,black] {$\cI$} -- (0.,2);

\end{scope}

\draw[thick, ->] (2.5,0) -- (4.5,0);
 \node at (3.5,0.3) {Folding};

\begin{scope}[shift={(7.,1)}, rotate=-90]
    
    \fill[top color=white!70!teal, bottom color=white!100!teal] (-1, -1.5) -- (0,0) -- (2,0) -- (1, -1.5) -- cycle;
        \draw (-1, -1.5) -- (0,0) -- (2,0) -- (1, -1.5);

    \fill[top color=white!70!teal, bottom color=white!100!teal] (0,0) -- (1,-2) -- (3,-2) -- (2,0) -- cycle;
      \draw (1, -2) -- (0,0) -- (2,0) -- (3, -2);
    
    \draw[red, line width=1.5] (0,0) -- (2,0); 
    \node at (2,0.25) {$B_{\cI}$};

    \node[left] at (-1.,-0.2) {\scriptsize CFT$_L$};
    \node[right] at (1.5,-1.5) {\scriptsize $ \overline{\text{CFT}}_R$};
    
      \draw[<->, thick, red!70!black, bend left=-35] (-0.5,-0.9) to (0.5,-1.2);

    \end{scope}

\end{tikzpicture}

    \caption{The folding trick maps an interface $\cI$ between $CFT_L$ and $CFT_R$ to a boundary condition $B_\cI$ for the folded double $CFT_L \times \overline{CFT}_R$.}
    \label{fig: folding}
\end{figure}

The mechanism proceeds as follows. Consider an interface $\cI$ between $\mathrm{CFT}_L$ and $\mathrm{CFT}_R$, whose symmetries\footnote{We define symmetries as the set of topological defects of each CFT, including both conventional (flavor) symmetries and discrete or generalized non-invertible symmetries.} we denote by $\cC_L$ and $\cC_R$, respectively. The interface is conformal if it is invariant under the reduced conformal group $SO(d-1,2)$. In $d=2$, this amounts to:
\be
T_L -\overline{T}_L = T_R - \overline{T}_R \quad \text{at the interface.}
\ee
Exploiting the residual conformal symmetry, we map the interface to a conformal boundary condition $B_\cI$ for the folded theory $\mathrm{CFT}_L \times \overline{\mathrm{CFT}}_R$, as illustrated in Figure~\ref{fig: folding}. This procedure is known as the folding trick.

We show that, in several examples, the symmetry of the folded theory is enhanced:
\be
\cC_{L \times \overline{R}} \supset \cC_L \times \overline{\cC_R} \, .
\ee
In particular, it may include new \emph{continuous} symmetries. We refer to these enhanced symmetries in the folded theory as \emph{phantom symmetries}. A key feature is that these symmetries are generically non-invertible. The spontaneous breaking of continuous non-invertible symmetries, recently analyzed in \cite{Damia:2023gtc}, leads to moduli spaces with orbifold geometry. Similarly, their explicit breaking by a boundary condition produces orbifold boundary conformal manifolds. 

We argue in Section \ref{sec:generic arguments} that, upon unfolding the boundary CFT (BCFT), although the continuous symmetry is generically lost, the boundary conformal manifold maps to a nontrivial -- typically non-homogeneous -- interface conformal manifold in the unfolded theory.

The introduction of a symmetry-based language enables the derivation of several properties of interfaces related by a broken phantom symmetry, including their $g$-functions and reflection coefficients (Section~\ref{sec: reflection}). Interestingly, we show that, while the $g$-function remains constant on the interface conformal manifold, the reflection coefficient $\cR$ does not. This must be contrasted with conformal manifolds arising from the breaking of ordinary symmetries, where also $\cR$ is constant. In our case $\cR$ changes as a consequence of the fact that the phantom symmetry does not commute with the stress tensors $T_L, \, T_R$ of the single CFTs, but only with their sum $T_L + T_R$.

We also believe that the phantom symmetry approach could give strong constraints on other interesting properties of defects and interfaces, such as the defect central charge $c_{eff}$ \cite{Sakai:2008tt,Brehm:2015lja,Wen:2017smb,Karch:2023evr,Karch:2024udk} and their fusion \cite{Bachas:2007td,Kravchuk:2024qoh,Diatlyk:2024zkk} (for which one could use e.g. \cite{Bachas:2001vj}). We leave these questions for future work.

It is tempting to conjecture that the breaking of a (possibly phantom) continuous symmetry is the only way in which conformal manifolds of codimension-one defects can be constructed.\footnote{For higher codimension there are several known counterexamples if we consider supersymmetric line defects \cite{Drukker:2022pxk}.}

Our results demonstrate how symmetry, and its breaking, provide a framework for populating interesting corners of the space of defect CFTs (DCFTs), a landscape that remains rich and largely unexplored. Our endeavors form part of a larger picture, emerged predominantly during the last few years, in which (generalized) symmetries are used to constrain the physics of extended, non-topological, defects and excitations \cite{Konechny:2019wff,Aharony:2022ntz,Aharony:2022ntz,Choi:2023xjw,Copetti:2024rqj,Copetti:2024dcz,Copetti:2024onh, Antinucci:2024izg,Cordova:2024iti,Choi:2024tri,Gagliano:2025gwr,Brennan:2025acl,Ambrosino:2025myh}.

\section{phantom symmetry in the (1+1)d Ising CFT}
\begin{table}
    \centering 
    \begin{equation*}
    \begin{array}{|>{\centering\arraybackslash}m{2cm}
                  |>{\centering\arraybackslash}m{1.75cm}
                  |>{\centering\arraybackslash}m{1.75cm}
                  |>{\centering\arraybackslash}m{2cm}|} \hline
    \raisebox{1.5ex}{\text{Ising}}
        & \shortstack{\rule{0pt}{3ex} $(1,\theta)$ \\ $\theta \in [0,\pi/2)$} 
        & \shortstack{$(\epsilon,\theta)$ \\ $\theta \in [\pi/2,\pi)$} 
        & \shortstack{$(\sigma,\theta')$ \\ $\theta' \in [0,\pi/2]$} \\  
    \hline
    \shortstack{\rule{0pt}{3.5ex}$R=\sqrt{2}$ \\ $\text{Orbifold}$}  
        &  \multicolumn{2}{c|}{\raisebox{1.5ex}{$|D^+,\theta\rangle$}}
        & \raisebox{1.5ex}{$|N^+, \theta' \rangle $} \\ 
    \hline
    \text{Free fermion} 
        & \multicolumn{2}{c|}{\chi^* = e^{-2i\theta} \overline{\chi}} 
        & $\chi = e^{-2i\theta'} \overline{\chi}$ \\ 
    \hline
    g-\text{function} 
        & \multicolumn{2}{c|}{1} 
        & $\sqrt{2}$ \\ 
    \hline
    \(\mathcal{R}\) 
        & \multicolumn{2}{c|}{\cos^2(2\theta)} 
        & $\cos^2(2 \theta')$ \\ 
    \hline
    \text{Phantom sym.} 
        & \multicolumn{2}{c|}{L_\theta^{(m)}} 
        & $L_{\theta'}^{(w)}$ \\ 
    \hline  
    \end{array}
    \end{equation*}
    \caption{Several equivalent descriptions of conformal defects in Ising CFT, with their associated observables}
    \label{tab: Ising defects}
\end{table}
Conformal defects in the Ising CFT have been completely classified \cite{Oshikawa:1996dj,Bachas:2013ora} and organize into continuous families, labeled by a primary field and a point on an interval (see Table \ref{tab: Ising defects}).

Special points correspond to the topological Verlinde lines of the Ising CFT: $1 = (1,\pi/4)$, $\eta = (\epsilon,3\pi/4)$, and $\cN = (\sigma,\pi/4)$.
The first two families are continuously connected, as $(\epsilon,\pi/2) = (1,\pi/2)$.

All other non-topological defect lines arise by deforming the corresponding topological ones via the marginal operator $\epsilon$ on the defect.
This deformation is exactly marginal, generating a defect conformal manifold.

At first glance, the existence of these families is surprising, given that the Ising CFT possesses only discrete symmetries, generated by the $\bZ_2$ line $\eta$ and the Kramers-Wannier duality defect $\cN$ \cite{Frohlich:2004ef,Chang:2018iay}. Furthermore, as the families in Table \ref{tab: Ising defects} are parameterized by an orbifold $ S^1/\bZ_2$, they are incompatible with the breaking of a standard continuous symmetry.

Both phenomena admit a natural explanation upon folding, where the original DCFT is mapped to a BCFT. Folding two copies of the Ising model produces the $\bZ_2$ orbifold of the $c=1$ compact boson $\phi \sim \phi +2\pi$ at radius $R = \sqrt{2}$.\footnote{We adopt conventions where the self-dual radius is $R_{\text{sd}}=1$.} As first observed in \cite{Thorngren:2021yso}, this theory admits two continuous families of non-invertible topological defects:
\bea
L_\theta^{(m)} &= 2 \cos\left(-2\theta \int \frac{\star d\phi}{2\pi}\right), \\
L_{\theta'}^{(w)} &= 2 \cos\left( 2 i \theta' \int \frac{d\phi}{2\pi}\right) \ .
\eea
These lines satisfy the fusion rules\footnote{Fusion involving fixed points under the orbifold requires care. For instance, $L_0^{(m/w)}$ degenerates to $\unit + \eta$, where $\eta = \eta_L\eta_R$ is the quantum symmetry of the orbifold \cite{Vafa:1989ih}.}
\be
L_{\theta_1}^{(m/w)} \times L_{\theta_2}^{(m/w)} = L_{\theta_1+\theta_2}^{(m/w)} + L_{\theta_1-\theta_2}^{(m/w)}\,, 
\ee
and are non-invertible. This type of continuous non-invertible symmetry is sometimes called \emph{cosine} non-invertible symmetry, and has been extensively studied in the literature \cite{Nguyen:2021yld, Thorngren:2021yso, Heidenreich:2021xpr, Antinucci:2022eat, Damia:2023gtc}. 

Dirichlet and Neumann boundary conditions of the free boson, upon orbifolding, are acted upon by $L_\theta^{(m)}$ and $L_{\theta'}^{(w)}$, respectively, reproducing the Ising conformal defects. Notably, $L_\theta^{(m)}$ and $L_{\theta'}^{(w)}$ exhibit a mixed 't Hooft anomaly, preventing the existence of a boundary state invariant under both symmetries \cite{Jensen:2017eof,Thorngren:2020yht}.

An alternative perspective on $L_\theta^{(m)}$ and $L_{\theta'}^{(w)}$ will facilitate the generalisations. Consider the spectrum of the Ising CFT in the $\bZ_2$-twisted Hilbert space $\cH_\eta$. It consists of three primary fields:
\be
\varphi_{\frac{1}{2},0}\,, \qquad \overline{\varphi}_{0,\frac{1}{2}}\,, \qquad \mu_{\frac{1}{16},\frac{1}{16}}\,.
\ee
Here, $\varphi$ and $\overline{\varphi}$ denote left- and right-moving Majorana-Weyl fermions. The exactly marginal deformation $\epsilon$ appears in the OPE $
\varphi \times \overline{\varphi} \sim \epsilon $.
In the folded theory, the fermions give rise to two currents:
$j = \varphi_L \varphi_R\,, \ \overline{j} = \overline{\varphi}_L \overline{\varphi}_R\ $
living in the twisted sector of the diagonal $\mathbb{Z}_2$ symmetry $\eta_L \eta_R$.\footnote{Subscripts $L$ and $R$ refer to the two copies of the Ising model, these are not to be confused with chiral/anti-chiral field, which we distinguish by a bar.} From these, vector and axial currents are constructed: $j_V = (j + \overline{j})/2$, $j_A = (j - \overline{j})/2$.

Although conserved, these currents are non-local. Thus, to build the symmetry generators $L_\theta^{(m)}$ and $L_\theta^{(w)}$, one must deform the defect $1 + \eta_L \eta_R$ rather than the identity, by inserting these currents:
\bea 
&L_\theta^{(m)} = (1+\eta_L \eta_R) \exp\left( i \theta \int \star j_V \right) \, , \\
&L_{\theta'}^{(w)} = (1+\eta_L \eta_R) \exp\left( i \theta' \int \star j_A \right)  \, .
 \eea
This procedure defines a valid deformation of a symmetry-reflecting topological defect, as recently formalized in \cite{Antinucci:2024izg} (see also \cite{Popov:2025cha}). The protected scaling dimensions of $j$ and $\overline{j}$ ensure exact marginality, while the deformed defect remains topological due to current conservation. 

The emergence of additional global symmetries in the folded theory is also manifest in the fermionic description \cite{jordan1993paulische,Coleman:1974bu,Karch:2019lnn}. The folded theory corresponds to a single Dirac fermion, which has a $U(1)_V \times U(1)_A$ symmetry, absent in the unfolded theory of the single Majorana fermion.

This picture also elucidates the relation between the tilt operator for the phantom symmetry and the marginal deformation. Upon folding, the Majorana spinor $(\varphi_R, \overline{\varphi}_R)$ maps to $(\overline{\varphi}_R, \varphi_R)$. Defining a Dirac fermion $\psi = (\varphi_L + i\overline{\varphi}_R,\, \overline{\varphi}_L + i\varphi_R) = (\chi, \overline{\chi})$. Bosonic boundary conditions are mapped to fermionic ones as in Table \ref{tab: Ising defects} \cite{Oshikawa:1996dj,Quella:2006de,Bachas:2013ora}.

The tilt $j_V^\perp = \frac{1}{2}\overline{\psi}\gamma^1\psi = \varphi_L\overline{\varphi}_R + \overline{\varphi}_L\varphi_R$ on the $D_\theta$ boundary, reduces to $\csc(2\theta)\epsilon$, the same happens for $j_A^\perp = \csc(2\theta')\epsilon$ on the $N_{\theta'}$ boundary.\footnote{Notice that fully reflecting interfaces are realized at $\theta=0,\,\pi/2,\,\pi$, corresponding to infinite marginal coupling. This aligns to some extent with the conclusion reached in \cite{Popov:2025cha} regarding relevant pinning-field deformations. }
Several properties of these conformal defects \cite{Quella:2006de,Bachas:2013ora} are also summarized in Table \ref{tab: Ising defects}. Notice in particular that, while their $g$-function is a constant along the conformal manifold--and corresponds to the quantum dimension of the underlying topological line-- the reflection coefficient changes. This is in contrast with the known fact \cite{Quella:2006de}, that $\cR$ is invariant under fusion with L and R topological defects. We will explain in detail in Section \ref{sec: reflection} how this property is intimately related with the breaking of a phantom symmetry.

\section{Phantom Symmetry and Defect Conformal Manifolds}\label{sec:generic arguments}
We now present the general mechanisms for constructing these defect conformal manifolds, extending the picture discussed for the Ising CFT.

Throughout this work, we will focus on ``phantom symmetries'' stemming from non-local currents \( j_A \) belonging to the twisted Hilbert spaces \( \mathcal{H}_{\eta_A} \), where \( \eta_A \) are invertible Abelian topological lines of the folded theory: \( \eta_A \times \eta_B = \eta_B \times \eta_A \).
Extensions to non-Abelian and even non-invertible symmetries \( \eta_A \) are likely possible, but we do not pursue them here.

The construction relies on three ingredients:
\begin{enumerate}
    \item A set of anomaly-free symmetry generators \( \eta_A \) in \(\mathrm{CFT}_L \times \mathrm{CFT}_R\). We will assume that each \( \eta_A \) factorizes into topological lines of the single CFTs as \( \eta_A = \eta_{A,L} \, \eta_{A,R} \).
    \item The existence of current operators \( j_A \in \mathcal{H}_{\eta_A} \), neutral under all symmetries \( \eta_A \).
    Phantom currents must take the form \( j_A = \psi_{L,A} \psi_{R,A} \), ensuring that the corresponding symmetries are genuinely absent in the single copy CFTs.
    \item A topological line \( \mathcal{A} \), satisfying \( \mathcal{A} \times \eta_A = \mathcal{A} \) for all \( \eta_A \). Given our assumptions we have
    \begin{equation}
    \mathcal{A} = \bigoplus_{\eta \in \Gamma} \eta \,,
    \end{equation}
    where \( \Gamma \) is the group generated by the \( \eta_A \).
\end{enumerate}
The requirement that the \( \eta_A \) be anomaly-free follows from the spin selection rule \cite{Chang:2018iay,Lin:2021udi} for \( \mathcal{H}_{\eta_A} \), which constrains the allowed spins in $\cH_{\eta_A}$ to lie in 
\begin{equation}
s_A \in \frac{\alpha _A}{N_A^2}+ \frac{1}{N_A}\bZ  \,,
\end{equation}
where \( \alpha_A \in H^3(\mathbb{Z}_{N_A}, U(1)) =\bZ_{N_A} \) is the anomaly class for the \( \mathbb{Z}_{N_A} \) symmetry generated by \( \eta_A \). It is straightforward to show that all bosonic operators in \( \mathcal{H}_{\eta_A} \) must furthermore be neutral under the \( \mathbb{Z}_{N_A} \) symmetry.

Strong constraints on the allowed current algebras for given $\eta_A$ follow from the OPE:
\begin{equation} \label{eq: phantomope}
j_A(z) j_B(w) \sim \frac{k_{AB}}{(z-w)^2} \, \mathbf{1} + i {f^C_{AB}}\, \frac{j_C(w)}{z-w}  \, .
\end{equation}
which should be thought of as graded over $\Gamma$. By assuming that the currents $j_A$ are attached to simple invertible lines $\eta_A$, the OPE determines the fusion of the $\eta_A$, and hence the underlying finite symmetry group $\Gamma$.

Let us give two examples illustrating these constraints.
\begin{itemize}
    \item[(a)] The OPE of a single $\fu(1)$ current $j(z)$ requires the latter to either be genuine or to live in the twisted sector of a $\bZ_2$ line $\eta^2=1$. 
    Multiple $\fu(1)$ currents, on the other hand, allow for more possibilities, depending on which basis of currents lives at the end of the simple topological lines.
    \item[(b)] Three currents \( j_{1,2,3} \) can close into an \( SU(2)_k \) current algebra. The allowed simple lines which can host $j_i$ in their twisted sector depends on the basis of the $\mathfrak{su}(2)$ algebra.
    If \( j_i \) are associated with the standard Pauli basis of \( \mathfrak{su}(2) \), then the lines \( \eta_1, \eta_2, \eta_3 \) give rise to the three \( \mathbb{Z}_2 \) subgroups of \( \mathbb{Z}_2 \times \mathbb{Z}_2 \). If instead the basis is (\( \sigma_3, \sigma^\pm \)), then \( j_3 \) must be a genuine current, while \( j_\pm \) are attached to lines \( \eta_\pm \) fulfilling \( \eta_+ \eta_- = 1 \).   
\end{itemize}
Let us emphasize that, when working with a phantom symmetry, we cannot arbitrarily change the basis of the Lie algebra: arbitrary linear combinations of operators from different twisted Hilbert spaces $\cH_{\eta^A}$ are not meaningful, but only combinations with positive integer coefficients. The assumption of simplicity of the lines $\eta_A$ to which $j_A$ are attached is not a restriction, as that basis is always meaningful. It is however imporant to notice that the same Lie algebra can support different finite symmetry groups $\Gamma$, depending on which basis of the phantom current\eqref{eq: phantomope} is attached to simple lines.\footnote{For instance while a single $\fu(1)$ currents is only compatible with $\Gamma=\bZ_2$ (or trivial), multiple $\fu(1)$ currents allow for more possibilities depending on which basis of currents live at the end of simple lines. An example is the symmetric product orbifold $S_N$ of a CFT with $U(1)^N$ symmetry. Focusing on $\bZ_N\subset S_N$ that maps $j_a\mapsto j_{a+1}$, the currents in the twisted sector of the simple lines $\eta ^k$ generating the dual $\bZ_N^\vee$ symmetry are not $j_a$, but $\widetilde{j}_k:=\sum _a e^{2\pi i \frac{ak}{N}} j_a$, whose OPEs are $\widetilde{j}_{k_1}(z)\widetilde{j}_{k_2}(w)\sim \frac{\delta_{k_1,-k_2}}{(z-w)^2} $.}

Under the conditions above, a topological line operator \( U_{\varepsilon} \) can be defined as
\begin{equation}
U_{\varepsilon}[\gamma] = \mathcal{A}[\gamma] \exp\left( i \varepsilon^A \int_{\gamma} \star j_A \right) \,, \ \ \ \begin{tikzpicture}[baseline={(0,1)}]
   \draw[thick, blue] (0,0) node[below] {$\cA[\gamma]$} -- (0,2); 
   \draw[dashed, very thick] (-1,1) -- (0,1);  
    \draw[fill=black] (0,1) circle (0.05);
    \draw[fill=black] (-1,1)  node[left] {$
    j$} circle (0.05);
    \node[left] at (-0.35,1.2) {\scriptsize $\eta$};
\end{tikzpicture}
\end{equation}
where \( \varepsilon^A \) are continuous parameters, and contractions over group indices are implicit. This construction generates a continuous symmetry of the folded theory, that is non-invertible whenever $\cA[\gamma]$ is non-trivial.

An alternative but useful point of view is to gauge the anomaly-free symmetry \( \eta^A \). In the orbifolded theory, the currents \( j^A \) become genuine and form a group $G_{\text{orb}}$. The orbifold theory has a dual $\widehat \Gamma$ symmetry, under which the currents are charged. The symmetry operator takes the form
\begin{equation}
U_{\varepsilon}[\gamma] = \bigoplus_{r \in \widehat \Gamma}  \exp\left( i \varepsilon^A \int_\gamma r(j^A) \right) \,,
\end{equation}
making its non-invertible nature manifest (see e.g. \cite{Thorngren:2021yso,Bhardwaj:2022yxj,Antinucci:2022eat,Damia:2023gtc} for discussions on this construction). Clearly, the parameters $\varepsilon^A$ belong to an orbifold $G_{\text{orb}}/\widehat \Gamma$.

A generic boundary condition will transform non-trivially under the $U_\varepsilon$ action, generating a conformal manifold. As anticipated in the introduction, the exactly marginal deformation is the \emph{tilt operator}, obtained by pulling back to the boundary the perpendicular component of the conserved current:
\be
t(x) = j \vert_{B_{\cI}} \, . 
\ee
In terms of the symmetry generator, the situation is more subtle because of the non-invertibility of the phantom symmetry. A reference boundary condition \( B \) generally transforms under a non-invertible (NIM) representation of \( U_\varepsilon \):
\(
U_\varepsilon \, |B\rangle = \sum_{B'} {n_{\varepsilon}}_B^{B'} \, |B'\rangle \, \)
so that acting with \( U_\varepsilon \) generally produces a non-simple boundary condition. In practice, the non-simplicity is mild and simple components can be extracted from the knowledge of the fusion algebra of \( U_\varepsilon \). Notice that this does not preclude the definition of an infinitesimal deformation by the tilt operator, but it influences the global topology of the conformal manifold.

When studying defects, rather than interfaces, it is particularly simple to consider reference boundary conditions \( |\mathcal{L}\rangle \) arising from folding along a topological line \( \mathcal{L} \). In this case, the boundary condition relates left- and right-moving operators via
\begin{equation}
\mathcal{O}_L = \mathcal{L}(\overline{\mathcal{O}}_R) \,.
\end{equation}
From the definition of the conserved current \( j = \psi_L \psi_R \), we find that the tilt operator, once the BCFT is unfolded, is mapped to the marginal operator
\begin{equation}
\mathcal{O}_{\Delta = 1} \sim \psi_L \times \mathcal{L}(\overline{ \psi_R}) \,,
\end{equation}
which remains exactly marginal when inserted along a line, by virtue of the inherited symmetry from the folded theory. 
For interfaces between different CFTs, however, the exactly marginal operator typically has no simple description purely in terms of the bulk fields.

\section{g-function and reflection coefficient}\label{sec: reflection}
The conformal manifold approach also allows for a simple computation of several interesting quantities characterizing the defect. Here we give two examples: the $g$-function \cite{Affleck:1991tk} and the reflection coefficient $\cR$ \cite{Quella:2006de,Bachas:2013ora,Meineri:2019ycm}.
\vspace{0em}
\paragraph*{\textbf{g-function}.} For a conformal line defect $\calD$, the $g$-function is defined as the (scheme independent) finite part of the normalized defect free energy on the circle:
\be
F_{\calD} = \log(g) + ... \, .
\ee
This is a monotonic quantity along defect RG \cite{Affleck:1991tk,Friedan:2003yc,Casini:2016fgb,Cuomo:2021rkm} and thus of great importance in constraining the IR behavior of such systems.
As expected, even for a defect conformal manifold generated by a phantom symmetry the $g$-function is constant. This can be easily derived re-expressing the free-energy as a disk partition function of the folded theory ($H=H_L + H_R)$:
\be
\frac{d}{d\varepsilon}g(\varepsilon) \propto \langle 0 | q^{H} \int \star j | B_{\varepsilon} \rangle = 0 \, .
\ee
Where we have used $\int \star j |0\rangle = 0$. If a defect conformal manifold is generated by deforming a topological line $\cL$, then $g = \langle \cL \rangle$ is the quantum dimension of $\cL$.
\vspace{1em}
\paragraph*{\textbf{Reflection coefficient}.} Another interesting set of observables for a defect CFT are the reflection and transmission coefficients $\cR,\cT=1-\cR$ \cite{Quella:2006de}. They can be defined as the fraction of reflected/transmitted energy in the scattering process involving the defect and are computed by the 2-point function of the stress tensors on the two sides of the defect \cite{Meineri:2019ycm}. 
 In particular $\cR=0$ signals the presence of a topological defect, while $\cR=1$ corresponds factorized ones \cite{Popov:2025cha}.
 
Defects related by a symmetry action share the same reflection coefficient \cite{Quella:2006de}. This is not the case for defects related by a phantom symmetry, as there is no corresponding global symmetry in the unfolded CFT. In fact, in folded theory the topological lines $U_{\varepsilon}$ only commute with the total stress tensor $T = T_L + T_R$. 

Consider, for simplicity, a $U(1)$ phantom symmetry generated by a single current $j=\psi_L\psi_R$ in the twisted sector of a $\bZ_2$ line $\eta$. The folded theory has a genuine (2,0) primary field \cite{Quella:2006de} 
\be
W^+ \propto c_R T_L - c_L T_R \, \ .
\ee
We can also construct a second, ``$\eta$-twisted" $(2,0)$ primary
\be
W^- \propto h_R \partial \psi_L \psi_R - h_L \psi_L \partial \psi_R \, .
\ee
In the Supplemental Material we show that these operators form a closed multiplet under the $U(1)$ phantom symmetry, which reads, after unit normalization,
\be
[Q,W^+] = - i n W^- \, , \ \ \ [Q,W^-] = i n W^{+} \, ,
\ee
$n\in \bZ$ being the the phantom symmetry charge. For our purposes it is most convenient to compute the reflection and transmission coefficients as defined in \cite{Quella:2006de}. They are determined uniquely by the 2-point function of $W^+$ in the folded theory 
\be
\langle 0 | W^{+}(z) \overline{W}^{+}(w) | B \rangle =  \frac{\omega^+_B/2}{(z-w)^4} \, ,
\ee
in terms of which \(\cR = \left(c_L^2 + c_R^2 + 2 c_L c_R \omega_B^+\right)/(c_L + c_R)^2 \).
We find it natural to also introduce the coefficient $\omega_B^{-}$ via \( \langle 0 | W^{-}(z) \overline{W}^{-}(w) | B \rangle =  \omega^-_B/\left(2(z-w)^4\right) \, \).
The multiplet structure allows to determine $\omega^{+}_B$ along the conformal manifold in terms of the coefficients $\omega^{\pm}_B$ computed at a reference boundary condition $|B_\epsilon\rangle = U_\varepsilon |B_0\rangle$ (see the Supplemental material for details)
\be
\omega_B^+(\epsilon) = \cos^2(n \epsilon) \omega_{B_0}^+ + a \sin^2(n \epsilon) \omega_{B_0}^- \, ,
\ee
where $a=\pm$ refers to a deformation by the vector/axial phantom symmetries respectively. Choosing either a transmissive or factorized boundary condition we can explicitly compute the coefficients $\omega_{B_0}^\pm$:
\begin{itemize}
    \item[(a)] If $CFT_L=CFT_R$ we can take $|B_0\rangle = |\cL \rangle$ to be defined by folding the theory over a topological line $\cL$. This is a transmissive boundary condition, hence $\omega_{\cL}^+=-1$. Moreover $\omega_{\cL}^-$ can be computed by unfolding the theory around $\cL$. Denoting by $q^{\cL}_{\psi_{L,R}}$ the eigenvalues of $\cL$ on $\psi_{L/R}$ we have $\omega_{\cL}^- = q^{\cL}_{\psi_L} q^{\cL}_{\psi_R} $, so that
    \be \label{eq:Rtop}
    \cR(\epsilon) = \frac{1 \pm q^{\cL}_{\psi_L} q^{\cL}_{\psi_R}}{2} \sin^2(2\epsilon) \, .
    \ee
    Notice that $|q^{\cL}_{\psi_L}|\leq 1$, so the reflection coefficient is always positive and strictly smaller than $1$ if $q^{\cL}_{\psi_L} q^{\cL}_{\psi_R}\neq 1$.
    
    \item[(b)] If the boundary condition is factorized, $|B_0\rangle = |B_L\rangle |B_R \rangle $, we have $\omega_{B_0}^+ = 1$ instead.  As $\psi_L$ and $\psi_R$ are chiral fields, their 2-point functions in the presence of a boundary take the form
    \(
 \langle 0 | \psi_{L/R}(z) \overline{\psi}_{L/R} (\overline{w}) |B_0\rangle = \beta_{L/R}/(z-w)^{2h_{L/R}} \). Hence $\omega_{B_0}^- = \beta_L \beta_R/k$, where $k$ is the level of the $U(1)$ current algebra generated by the phantom current. The reflection coefficient reads
    \be \label{eq:Rfact}
    \cR(\epsilon) = 1 - \frac{2 c_L c_R}{(c_L + c_R)^2}\left(1\mp \frac{\beta_L \beta_R}{k} \right) \sin^2(n \varepsilon) \, .
    \ee
    The minimum reflection coefficient along the conformal manifold is $\cR^*=1 - 2(1+|\omega_{B_0}^-|)c_L c_R / (c_L + c_R)^2$. As $|\omega_{B_0}^-|<1$ unless $c_L=c_R$ --in which case $|\omega_{B_0}^-|=1$ for topological interfaces-- the reflection coefficient is positive definite.
\end{itemize}
As a check, consider the conformal defects in the Ising CFT. Here $\psi_{L/R}$ are the chiral Majorana fields $\varphi_{L/R}$. Starting at the topological points with reference boundaries $|\unit\rangle$ or $|\eta\rangle$ the composite charge $q^{\cL}_{\varphi_L} q^{\cL}_{\varphi_R}$ is trivial and we act with the vector current, so $\cR(\epsilon) = \sin^2(2\varepsilon)$, consistent with the results in the literature \cite{Bachas:2013ora}. If we consider $|\cN\rangle$ instead, we have $q^{\cN}_{\varphi_L} q^{\cN}_{\varphi_R} = -1$ and we act with the axial current. Again the result is $\cR(\varepsilon')= \sin^2(2 \varepsilon')$. We can also start at the factorized points, using the explicit Dirichlet boundary conditions of Table \ref{tab: Ising defects} for the Majorana fermions we find, for all factorised points, $\beta_L\beta_R=-1$ so $\omega_{B_0}^{-}=-1$ and \(
    \cR(\varepsilon) = \cos^{2}(2\varepsilon)=\sin^{2}\left(2\left(\varepsilon-\frac{\pi}{4}\right)\right)\, \), in agreement with the previous answer.
Similarly starting from the Neumann boundary conditions we have $\beta_L\beta_R=1$ at all factorized points, and, since here the phantom symmetry is the axial one, we recover the same result $\cR(\varepsilon) = \cos^{2}(2\varepsilon)$ as expected.

\section{Examples}
We conclude with listing several examples in which phantom symmetries appear, apart from the Ising CFT.
\vspace{0.5em}
\paragraph*{ $\mathbf{c =1}$ \textbf{theories.} } We begin by analyzing the two-dimensional $c=1$ conformal field theory of a free compact boson $\phi$ at radius $R$. Its vertex operators \( V_{n,w} = :e^{in\phi + i w \widetilde \phi}: \ n,w \in \bZ \) have conformal dimensions 
\(
 h=\frac{1}{4}\left(\frac{n}{R}+wR\right)^2\, , \ \bar{h}=\frac{1}{4}\left(\frac{n}{R}-wR\right)^2 \) in our conventions.
 Let us consider when the identity defect admits a marginal deformation. Scalar solutions with $h=\bar h = 1/2$ exist only for rational $R^2$ and are:
\be\label{eq: c01 marginals}
V_{\pm 2n, 0} \ \text{at } R = \sqrt{2}n, \quad V_{0, \pm n} \ \text{at } R = \frac{\sqrt{2}}{n}\,,
\ee
and their $T$-duals. At $R = \sqrt{2}$ both sets of operators are allowed.

We now explore whether these deformations are exactly marginal, and thus span a defect conformal manifold. We first observe that at these specific values of the radius, the theory also contains non-genuine fermionic operators of dimension $(h=1/2,\bar h = 0)$, $(h=0,\bar h = 1/2)$, summarized in Table \ref{tab: twisted boson}.
\begin{table}
\centering
\begin{equation*}
\begin{array}{|c|c|c|}
\hline
{}  & R = \sqrt{2}n & R = \sqrt{2}/n \\
\hline
\left( \tfrac{1}{2}, \tfrac{1}{2} \right) & V_{\pm 2n, 0} & V_{0, \pm n} \\ \hline
\left( \tfrac{1}{2}, 0 \right) 
    &  \raisebox{0.5ex}{$V_{n,\,\frac{1}{2n}} ,\; V_{-n,\,-\frac{1}{2n}}$} 
    &  \raisebox{0.5ex}{$V_{\frac{1}{n},\,\frac{n}{2}} ,\; V_{-\frac{1}{n},\,-\frac{n}{2}}$} \\
\hline
\left( 0, \tfrac{1}{2} \right) 
    &  \raisebox{0.5ex}{$V_{-n,\,\frac{1}{2n}} ,\; V_{n,\,-\frac{1}{2n}} $}
    &  \raisebox{0.5ex}{$V_{\frac{1}{n},\,-\frac{n}{2}} ,\; V_{-\frac{1}{n},\,\frac{n}{2}}$} \\
\hline
 \raisebox{1.75ex}{$\eta$} 
    & \raisebox{1.75ex}{$ \left( 0,\,\tfrac{1}{2n} \right) ,\; \left( 0,\,-\tfrac{1}{2n} \right)$ }
    & \shortstack{ \rule{0pt}{2.5ex}
       $\left( \tfrac{1}{n},\,0 \right)$ ,\; $\left( -\tfrac{1}{n},\,0 \right)$ \ $\text{if } n \text{ even} $\\[0.3em]
 $\left( \tfrac{1}{n},\,\tfrac{1}{2} \right)$ ,\; $\left( -\tfrac{1}{n},\,\tfrac{1}{2} \right)$  $\text{if } n \ \text{odd}$
      } \\
\hline
\end{array}
\end{equation*}
    \caption{Marginal deformations and twisted chiral operators giving rise to phantom symmetry for the free boson. The twisted sectors in which the chiral operators lie are denoted by $(\alpha, \beta)$, corresponding to the generator $(2\pi\alpha, 2\pi \beta)$ of the $U(1)_m \times U(1)_w$ symmetry.}
    \label{tab: twisted boson}
\end{table}

The marginal operators in \eqref{eq: c01 marginals} arise in the OPE of the chiral and antichiral operators.

Upon folding, the chiral operators generate $4$ non-genuine holomorphic (anti-holomorphic) currents from their products. Therefore the folded theory has an enhanced phantom symmetry at these radii. Decomposing the scalar fields $\phi$ and $\widetilde \phi$ into (anti)chiral components $X$ and $\overline{X}$ as
\(
\phi = \frac{1}{\sqrt{2}R}(X + \overline{X})\,, \ \widetilde{\phi} = \frac{R}{\sqrt{2}}(X - \overline{X}) \) the (anti)chiral operators are simply $e^{\pm i X_{L/R}}$, $e^{\pm i \overline{X}_{L/R}}$. 
Denoting $X = X_L + X_R$ we have the OPE:
\begin{equation}
e^{+ i X}(z) \times e^{- i X}(w) = \frac{1}{(z-w)^2} + \frac{i}{z-w} \, \partial X(w) + ...\,,
\end{equation}
with an analogous expression holding for the anti-chiral fields and the anti-diagonal combination $X'= X_L - X_R$, that have regular OPEs with $e^{\pm i X}$. We conclude that the folded theory has a phantom symmetry with algebra $\mathfrak{su}(2) \times \mathfrak{su}(2) \cong \mathfrak{so}(4)$. 
In accordance with the general arguments presented in Section \ref{sec:generic arguments}, the non-Cartan currents appear in the twisted sector of mutually inverse group elements, which form $\Gamma = \bZ_{2n}$ or $\bZ_2 \times \bZ_n$ depending on the radius.
The reason behind this symmetry enhancement is manifest in the fermionic description of the theory. At $R = \sqrt{2}$, the $c=1$ compact boson is the bosonization of a free Dirac fermion. Thus, after folding, we find a theory of 2 free Dirac fermions, whose global symmetry is enhanced to \( 
O(4)_L \times O(4)_R \supset U(1) ^ 4 \).
A generic boundary condition will break the global symmetry to an $O(4)$ subgroup, so the defect conformal manifold will be spanned by $\text{dim}(O(4)) = 6$ generators. The $U(1)$ symmetries that span the Cartan subgroup of $O(4)$ remain exact symmetries in the unfolded theory. The remaining 4 generators correspond to the marginal deformations described earlier, and show up as phantom symmetries in the bosonic description. 
The values $R=\sqrt{2}n, \sqrt{2}/n$ are obtained by gauging $\bZ_n$ subgroups of the momentum or winding symmetries. This procedure breaks the $O(4)$ symmetry down to its $SU(2) \times U(1)$ subgroup, thereby eliminating 2 of the 6 marginal deformations of the $R = \sqrt{2}$ theory.
An analogue phantom symmetry algebra can be defined at any radius $R= \sqrt{2}p/q$, with the chiral operators being $V_{\pm p/q,\pm q/2p}$. At these radii there are however no marginal deformations of the identity defect and a phantom conformal manifold will only appear upon deformation of suitably symmetry-reflecting \cite{Antinucci:2024izg} conformal/topological defects.
Clearly the pattern of phantom symmetries in $c=1$ theories is much richer, especially if we consider interfaces between $c=1$ theories at different points of the conformal manifold. It would be interesting to revisit the results of \cite{Bachas:2007td} from this perspective.
 \vspace{0.5em}
\paragraph*{\textbf{Interfaces between RCFTs.}} Another solvable set of examples are interfaces between RCFTs, for example $G_{k}$ and $G'_{k'}$ WZW models. Interfaces preserving some (or all) of the chiral algebra have been studied extensively \cite{Cardy:1989ir,Fuchs:2002cm,Quella:2002ct,Quella:2006de}. Our results will instead focus on symmetry breaking interfaces.  
Let us consider systems with a single Abelian phantom current $j$. We must then either have that $\eta_L$, $\eta_R$ are anomaly-free and $h_L=h_R=\frac{1}{2}$ or that both $\eta_L$ and $\eta_R$ are anomalous and $h_L=\frac{1}{4}$, $h_R=\frac{3}{4}$. It is simple to classify WZW models with such operators if $\psi_L$ and $\psi_R$ are chiral primaries. Taking the $\bZ_2$ to be a subgroup of the center of $G$, the only chiral operators is the $\bZ_2$ twisted sector are order two simple currents $(J,0)$ \cite{Schellekens:1990xy}.
This follows immediately from the twisted partition function of diagonal RCFTs $Z_\lambda = \sum_{\mu, \bar \mu} N_{\mu \bar{\mu}}^\lambda \, \chi_\mu \, \bar \chi _{\bar \mu}$ taking $\bar{\mu}=0$.
It is immediate to compute their conformal dimensions. Their list can be found in Table \ref{tab: WZW}.
\begin{table}
    \centering
    \begin{equation*}
\begin{array}{|c|c|} \hline 
h = 1/2  &SU(2)_2, \, SU(4)_1, \, \text{Spin}(n)_1 \, , \\ \hline
h = 1/4  &SU(2)_1 , \, \\ \hline
h = 3/4   &SU(2)_3, \, SU(6)_1 , \, \text{USp}(6)_1 , \, \text{Spin}(12)_1 , \, (\text{E}_7)_1 \\ \hline
\end{array}
\end{equation*}
    \caption{Relevant WZW models for constructing Abelian phantom symmetries by embedding $\eta$ in their common center.}
    \label{tab: WZW}
\end{table}
A phantom current will appear in interfaces between theories in the first row, and between theories in the second and third row.
In these examples, the operators $\psi_L$ and $\psi_R$ carry a nontrivial representations under the left and right chiral algebras. The chiral algebra of the WZW is thus completely broken on the conformal manifold, apart from a set of isolated points. The $G$ and $G'$ action on $j$ thus generate new phantom currents.

We can also understand when the theory has an phantom symmetry with $\mathfrak{su}(2)$ algebra. We will try to find chiral currents in the twisted sectors $\cH_{\eta^m}$ and $\cH_{\eta^{-m}}$ of the center symmetry. The simplest group that admits this structure is $\bZ_3$. Chiral operators with dimension $\frac{1}{3}$in the $\bZ_3$ twisted sector are present in $SU(3)_1$ only, while the dimension $\frac{2}{3}$ is present in $SU(3)_1$ and $(E_6)_1$. This gives rise to families of interfaces with $\mathfrak{su}(2)$ phantom symmetry between these models.

Finally we comment that one could use this list to construct examples where no global symmetry in the single CFT is present. For example consider the GKO cosets $G_k \times G_l/G_{k+l}$ \cite{GOODARD198588}. They can be thought of the IR of the theory in the numerator after gauging the diagonal vector symmetry.\footnote{More precisely, one must gauge $G/Z(G)$ to avoid a theory with multiple universes \cite{Hellerman:2006zs,Komargodski:2020mxz,Cordova:2023jip,Damia:2024kyt}.} Chiral simple currents $J$, $J'$ in the $\bZ_2$ twisted sectors are protected by the spin selection rule, but they must combine together to have trivial charge under the vector symmetry. In this way, one recovers, for example $SU(2)_1 \times SU(2)_1/SU(2)_2=$Ising. Moreover, the 3-state Potts CFT, $3\text{Potts}=SU(3)_1 \times SU(3)_1/SU(3)_2$, has a chiral operator of dimension $h=2/3$ in the $\bZ_3$ twisted sector of the theory. Therefore, we conclude that interfaces between 3-state Potts and $SU(3)_1$ have an $\mathfrak{su}(2)$ phantom symmetry. 

It is clear that these examples are only the tip of the iceberg of a large landscape of phantom symmetries. More examples can be uncovered by considering discrete symmetries which are not embedded in the center of $G \times G'$.
\vspace{0.5em}
\paragraph*{\textbf{Higher-dimensional examples.}} 
Phantom symmetries are also naturally present in $d>2$ free field theories, where they also predicts the existence of defect conformal manifolds. The simplest example is a single free real scalar that only has an $O(1)=\bZ_2$ symmetry, while the folded theory is a complex scalar with $O(2)$ symmetry. The identity defect is defined by the b.c. 
\begin{equation}
    \phi_L = \phi_R \ \ \ \ \  \  \ \ \text{and}\ \ \ \ \ \ \ \ \ \partial_\perp \phi_L = - \partial_\perp \phi_R
\end{equation}
of the folded theory, namely Dirichlet and Neumann for $\phi_L\mp \phi _R$ respectively. The phantom symmetry  corresponds to the $U(1)$ symmetry of a complex scalar, that contains $\bZ_2$ as a subgroup. This phantom symmetry is broken by the b.c. with tilt $t =2 i \phi_L \partial_\perp \phi_L$. The defect conformal manifold $\calD_\theta$ ($\theta =[0,\pi)$), which has an $S^1$ topology, also contains the $\bZ_2$ symmetry defect $\eta = \calD_{\pi/2}$, as in the Ising CFT. This follows from the fact that $\bZ_2$ becomes part of the phantom symmetry.

The general case of interfaces between $p$ scalars and $q$ scalars can be understood similarly. The two theories have symmetries $O(p)$ and $O(q)$ respectively, and there is a phantom $O(p+q)$ symmetry manifesting as the symmetry of the folded theory that has $p+q$ scalars. Boundary conformal manifolds of free $N$-component scalar theories have been studied in \cite{Herzog:2023dop}, where the authors noticed that their geometry is that of a coset (see also \cite{Kim:2025tvu} for codimension-1 defects). 
By choosing two integers $p',q'$ with $p'+q'=N=p+q$ and splitting the scalar fields into $\left\{ \phi_i, \, \phi_a \right\}$, $i=1,...,p'$, $a=1,...,q'$, we impose the boundary conditions \cite{Herzog:2023dop}:
\be
\Lambda_{a i} \partial_\perp \phi_a - \partial_\perp \phi_i = 0 \, , \qquad
\Lambda_{a i} \phi_i + \phi_a = 0 \ .
\ee 
Upon unfolding we get conformal interfaces between $p$ and $q$ free scalar theories with a coset geometry $O(N)/\left(O(p')\times O(q')\right)$ (also known as a Grassmannian).

Different choices of $p',q'$ such that $p'+q'=p+q$ label different connected components of the interface conformal manifold. Notice that, if $p=q$ (defect case), the component with $p'=q'=p$ has geometry $O(2p)/ \left(O(p)\times O(p)\right)$ and contains a submanifold $O(p)$ corresponding to the topological defects.

A similar result holds for interfaces between $p$ and $q$ free Majorana fermions, in which case the $O(p) \times O(q)$ symmetry is enhanced to $O(p+q)$ upon folding. The marginal deformation spanning the phantom conformal manifold, much as in the Ising case, can be described by the exactly marginal Majorana masses at the interface.

Let us close with some comments about possible interacting examples. Recall that the unitarity bounds for scalars, fermions and higher-spin primaries are, respectively \(
\Delta_\phi \geq \frac{d-2}{2} \, , \ \Delta_f \geq  \frac{d-1}{2} \, , \  \Delta_s \geq d+s-2 \), where the lower bounds are saturated by free fields and conserved (higher-spin) currents respectively.
To form a phantom current out of a composite of $\cO_L$ and $\cO_R$ we can then have only free scalars $J_\mu \sim \phi_L \partial_\mu \phi_R$ or free-fermion bi-linears $J_\mu \sim \bar\psi_L \gamma_\mu \psi_R$. All other composite operators of spin one must have $\Delta > d-1$ by the unitarity bounds. Thus phantom symmetries in interacting theories are intrinsic of (1+1)d\footnote{We thank Marco Meineri for pointing this simple argument out to us.}, that is the only case where the spin $s$ can take arbitrary values\footnote{Mathematically this is because $\pi_1\left(SO(2)\right)=\bZ$, while $\pi_1\left(SO(d)\right)=\bZ_2$ for any $d>2$.}. Possible ways to avoid this no-go theorem might include the presence of non-gauge invariant currents in the folded theory, such as in $\bR$ Maxwell theory in 4d.

\section*{Acknowledgments}
We thank Ho Tat Lam, Marco Meineri, Yifan Wang and Yunquin Zheng for illuminating discussions. The work of A.A. is supported by the UKRI Frontier Research Grant, underwriting the ERC Advanced Grant ``Generalized Symmetries in Quantum Field Theory and Quantum Gravity”. C.C.
is supported by STFC grant ST/X000761/1. The research of G.G. is funded through an
ARC advanced project, and further supported by IISN- Belgium (convention 4.4503.15). The research of A.A. and C.C. was supported in part by grant NSF PHY-2309135 to the Kavli Institute for Theoretical Physics (KITP). G.R. is supported by the Huawei Young Talent program at IHES. 

\newpage

 \setlength{\bibsep}{2pt plus 0.3ex}

 \bibliography{mybib.bib}

\newpage

\onecolumngrid

\begin{center}
    \textbf{\LARGE Supplemental Material}
\end{center}

\setcounter{section}{0}
\renewcommand\thesection{\Alph{section}}

\section{Phantom Symmetry action and reflection coefficients}\label{app:Wpm}
We consider the situation in which the folded theory has a single phantom current $j(z)=\psi_L(z)\psi_R(z)$ (and its antiholomorphic counterpart) in the twisted sector of a non-anomalous $\bZ_2$ symmetry. As we argue in the main text, deforming the condensate of the $\bZ_2$ symmetry by the integral of $j$ produces the phantom symmetry operators $U_{\epsilon}$. Our main objective is to compute the reflection coefficient as a function of $\epsilon$ on the defect conformal manifold. To do so, the first task is to find the action of the phantom symmetry operators on the stress tensors $T_L, T_R$. Knowing the action of $U_{\epsilon}$ on the stress tensors we can use the formulas of \cite{Quella:2006de, Meineri:2019ycm} to compute the reflection coefficients as we move on the defect conformal manifold. 

\subsection{Phantom Symmetry Multiplet}

This is most easily computed in the global variant in which the current is genuine, i.e. after gauging the $\bZ_2$ symmetry. We assume that the generator $\eta$ of the $\bZ_2$ symmetry is factorized $\eta=\eta_L\eta_R$, namely both $CFT_L$ and $CFT_R$ have a $\bZ_2$ symmetry hosting the operators $\psi_L$ and $\psi_R$ in their twisted sectors, so that in the folded theory the stress tensors $T_L$ and $T_R$ are invariant under the action of $\eta$ and survive the orbifold operation. The quantum $\widehat{\bZ}_2$ symmetry, generated by a line $\widehat{\eta}$, acts negating the current $j$, notice also that the operators $\psi_L$ and $\psi_R$ are not genuine if taken separately. In this global variant the phantom symmetry operator is simply
\begin{equation}
    U_\epsilon = e^{i\epsilon Q} + e^{-i\epsilon Q}\, , 
\end{equation}
where we introduced the charge operator
\begin{equation}
    Q = \int \frac{d z}{2\pi i} j(z)\, .
\end{equation}
The action of $U_\epsilon$ on the stress tensors follows from the commutation rules of the charge with both $T_L$ and $T_R$. Our first task is to identify the proper set of Virasoro primaries on which the symmetry acts. Out of $T_L$ and $T_R$ we construct the operators
\begin{equation}
    T= T_L+ T_R\, \qquad W^{+}= c_R T_L - c_L T_R
\end{equation}
the ratio of the definition of $W^{+}$ is that, from the OPE of $W^{+}$ with the total stress tensor $T$, one can see that is is a Virasoro primary of weight $(2,0)$ \cite{Quella:2006de}. From the OPE
\begin{equation}
    T_L(z) \, j(w) = \frac{h_L }{(z-w)^2} j(w) + \frac{1}{(z-w)} \partial \psi_L (w) \psi_R(w) \, , 
\end{equation}
and its analouge with $T_R$ we find that the operator
\begin{equation}
    W^{-}(z)= h_R \partial\psi_L(z) \psi_R(z) -  h_L \psi_R(z) \partial \psi_R(z) 
\end{equation}
is another Virasoro primary (with respect to the total stress tensor) of weights $(2,0)$. To compute the action of the phantom charge $Q$ on $W^{+}$ we start from the OPE
\begin{equation}
    W^{+}(z)j(\omega) =\frac{c_R h_L - c_L h_R}{(z-\omega)^2}j(\omega) + \frac{c_R\partial\psi_L(\omega)\psi_R(\omega)-c_L\psi_L(\omega)\partial\psi_R(\omega)}{z-\omega}
\end{equation}
expanding the RHS around $\omega=z$ we find
\begin{equation}
     W^{+}(z)j(\omega)\supset \frac{1}{z-\omega}\left(-(c_R h_L - c_L h_R)\partial j(z) + c_R\partial\psi_L(z)\psi_R(z)-c_L\psi_L(z)\partial\psi_R(z)\right) = \frac{c_{tot} W^{-}}{z-\omega}
\end{equation}
where we used $h_L+h_R=1$. Then, integrating over $\omega$ gives
\begin{equation}
    [Q, W^{+}] = -c_{tot}W^{-}\, .
\end{equation}
The remaining commutator to compute is $[Q,W^{-}]$, to do so we need the OPEs of $\psi_{L,R}$ with itself, in absence of other currents in both CFT$_L$ and CFT$_R$ these must take the form
\begin{equation}
    \begin{split}
        &\psi_L(z)\psi_L(w) = \frac{k_L}{(z-w)^{2 h_L}} + 2 h_L/c_L k_L (z-w)^{2 h_R} T_L(w) \, , \\ &
\psi_R(z)\psi_R(w) = \frac{k_R}{(z-w)^{2 h_R}} + 2 h_R/c_R k_R (z-w)^{2 h_L} T_R(w)\, .
    \end{split}
\end{equation}
Indeed, in absence of currents, the only other holomorphic operator of both CFT$_L$ or CFT$_R$ is the stress tensor (in particular $\psi_{L,R}$ itself cannot appear in this OPE because of the $\bZ_2$ symmetry selection rule), whose OPE coefficient is fixed in terms of that of the identity (see e.g. \cite{DiFrancesco:1997nk} (6.181)). From here it is easy to compute the commutator
\begin{equation}\label{eq:commQK}
    [Q,W^{-}] = -\frac{4 k_L k_R h_L h_R}{c_L c_R}  W^{+}\, . 
\end{equation}
Let us also notice that the level of the $U(1)$ chiral algebra generated by $j$ is determined as $k=k_L k_R$. 

\paragraph{Inclusion of global symmetry} Let us also show that these formulas continue to hold if (genuine) global currents $J^a_L$, $J^a_R$ are present in the single CFTs. 
$\psi_L$ is in a representation $R_L$ of $G_L$ , which must contain the identity in its tensor product:
\be
R_L \times R_L = 1 + ... 
\ee
The OPE of $\psi_{L/R}$ is modified to:
\be
\psi_L (z) \psi_L(w) = \psi_L(z)\psi_L(w) = \frac{k_L}{(z-w)^{2 h_L}} + \frac{r_{L \, a}}{(z-w)^{2 h_L -1}} J_L^a + 2 h_L/c_L k_L (z-w)^{2 h_R} T_R(w) + ...
\ee
where we suppress $R$ indices for the various OPE coefficients. Similar considerations are valid for $\psi_R$. The current(s) $j_d$ is now defined by contracting the $\psi_L \psi_R$ fields with a $R_L \times R_R$ tensor $d$.
One might wonder whether the presence of the currents might affect the phantom symmetry structure, as the operators $J_L^a J_R^{\alpha}$ are (2,0) primaries of the folded theory. 
This is not the case, consider the commutator $[Q_{d},j_d]=0$, using the OPE we find:
\be
[Q_{d}, j_d] = d \cdot \left( r_{L \, a} J^a_{L} + r_{R\, \alpha} J_R^\alpha \right) d = 0\, ,
\ee
which implies that $r_L$ and $r_R$ are anti-symmetric tensors. It is easy to see that this also kills any contributions to the commutator $[Q_d,K_d]$.

Thus, under these assumptions, the phantom charge acts as a rotation on the doublet $(W^{+},W^{-})$. This is the result one could have expected on general grounds: indeed since both $W^{-}$ and the current $j$ are charged under the quantum $\widehat{\bZ}_2$ symmetry, the operator $W^{-}$ cannot appear in the right-hand side of the commutator \eqref{eq:commQK}. Moreover only $W^{\pm}$ are Virasoro primaries, while $T_L, T_R$ and $T$ are not, then, since the charge $Q$ commutes with all Virasoro generators, it must map primaries into primaries and therefore the commutator $[Q,W^{-}]$ can only contain $W^{+}$. All in all the multiplet structure is 
\begin{equation}
    [Q, T]=0\, , \qquad [Q, W^{+}] =-c_{tot}W^{-} \, , \qquad [Q, W^{-}]= -\gamma W^{+}\, .
\end{equation}
where $\gamma=4 k_L k_R h_L h_R/(c_L c_R)$. These commutation relations completely determine the action of the phantom symmetry operators on the stress tensors. The charge can be represented as a $2\times 2$ matrix 
\begin{equation}
    \widehat{Q}= \begin{pmatrix}
0 &- c_{tot}  \\
-\gamma & 0 
\end{pmatrix}
\end{equation}
this matrix has eigenvalues $\pm\sqrt{c_{tot}\gamma}$, to ensure that the symmetry is an actual $U(1)$ we set $\gamma= n^{2}/c_{tot}$ so that
\begin{equation}
    n^2 = 4 k h_L h_R \frac{c_L + c_R}{c_L c_R} 
\end{equation}
and $n\in \bZ$. It is convenient to work in the normalizations
\begin{equation}
    W^{+} = -\sqrt{\frac{2}{c_L c_R (c_L + c_R)}}(c_R T_L - c_L T_R)\, , \qquad W^{-} = \frac{i}{\sqrt{2 h_L h_R k_L k_R}} (h_R \partial\psi_L(z) \psi_R(z) -  h_L \psi_L(z) \partial \psi_R(z) )
\end{equation}
these are chosen in such a way that the primary states are unit normalized, i.e. 
\begin{equation}
    W^{\pm}(z)W^{\pm}(\omega)=\frac{1}{(z-\omega)^4}+\dots
\end{equation}
and the action of the charge is simple
\begin{equation}
    [Q, W^{\pm}]= \mp i n W^{\mp}\, , \qquad   \widehat{Q}= \left(\begin{array}{cc}
0 & i n  \\
-i n & 0 
\end{array} \right)
\end{equation}
It is now straightforward to find the group action
\begin{equation}
    M(\epsilon)=e^{i\epsilon \widehat{Q}} = \begin{pmatrix}
\cos(n\epsilon)& - \sin(n \epsilon)
 \\ \sin(n \epsilon) & \cos(n\epsilon)
\end{pmatrix}\, .
\end{equation}
and the periodicity $\epsilon\sim \epsilon+ 2\pi$ is manifest. In terms of the operators $W^{\pm}$ we have
\begin{equation}\label{eq:Wpmtransf}
\begin{split}
     & e^{i\epsilon Q}W^{+}e^{-i\epsilon Q} = \cos(n\epsilon) W^{+} - \sin(n\epsilon) W^{-}\\
     & e^{i\epsilon Q}W^{-}e^{-i\epsilon Q} = \cos(n\epsilon) W^{-} +\sin(n\epsilon) W^{+}\, .
\end{split}
\end{equation}

\paragraph{An example.} As a simple check that this multiplet structure is correct consider the Ising case. The folded theory is the orbifold of the compact boson at $R= \sqrt{2}$ (in this conventions the self-dual radius is $R=1$) under the $\bZ_2$ charge conjugation symmetry. The global variant in which the phantom current is genuine is simply the compact boson before the orbifold. It is easy to see that this theory has $3$ operators with $h=2$ and $\overline{h}=0$, the first is the stress tensor $T=T_L+ T_R$ the remaining ones are the two vertex operators $V_{2,1}$ and $V_{-2,-1}$ which we arrange in the real combinations 
\begin{equation}
    W^+ =\left(V_{2,1}+ V_{-2,-1}\right)\, , \qquad W^{-} =i \left(V_{2,1}- V_{-2,-1}\right)
\end{equation}
the identifications with $W^{\pm}$ as defined above follows from the charges under the $\bZ_2$ charge conjugation symmetry. Depending on the reference boundary condition we choose, the phantom symmetry can be taken as either momentum or winding (better, their non-invertible $\bZ_2$-invariant counterparts), in any case it is easy to compute the commutators of the winding and momentum charges with these operators,
\begin{equation}
e^{i \epsilon Q} W^{\pm}e^{-i \epsilon Q} = \cos(n\epsilon) W^{\pm} \mp \sin(n \epsilon)W^{\mp}
\end{equation}
where $n$ is either the momentum or winding charge of $V_{2,1}$.

\subsection{Reflection and transmission coefficients along the conformal manifold}\label{subsec:reflcm}

This symmetry action can be used to extract the dependence of the relfection coefficient on the exactly marginal parameter of the defect conformal manifold. Given the simple action of the phantom symmetry on the doublet $(W^{+},W^-)$ the most direct way to access the reflection coefficient on the conformal manifold is to compute the expectation value 
\begin{equation}
    \langle 0 | W^{+}(z) \overline{W}^{+}(\overline{\omega})|b_{V,A}(\epsilon)\rangle
\end{equation}
where $|b_{V,A}(\epsilon)\rangle= U^{V,A}_{\epsilon}|b(0)\rangle$ and $|b(0)\rangle$ is a reference boundary condition. Here we have introduced the vector/axial phantom symmetries
\begin{equation}
    U_{\epsilon}^{V} = e^{i \epsilon(Q+ \overline{Q})}+ e^{-i \epsilon(Q+ \overline{Q})}\, \qquad U^A_{\epsilon}=e^{i \epsilon(Q- \overline{Q})}+ e^{-i \epsilon(Q- \overline{Q})}\, .
\end{equation}
According to \cite{Quella:2006de} we have, in our normalization,
\begin{equation}
    \langle 0 | W^{+}(z) \overline{W}^{+}(\overline{\omega})|b_{V,A}(\epsilon)\rangle = \frac{ \omega_{b_{V,A}}(\epsilon)}{(z-\omega)^4}
\end{equation}
where the parameter $\omega_b$ determines reflection and transmission coefficients as 
\begin{equation}
\begin{split}
      & \cR = \frac{c_L^2 +2 c_L c_R \omega_b + c_R^2}{c_{tot}^2}\\ & \cT = \frac{2 c_L c_R (1-\omega_b)}{c_{tot}^2}= 1-\cR\, , 
\end{split}
\end{equation}
in the notation of \cite{Meineri:2019ycm} one has $c_{LR}= c_L c_R (1-\omega_b)$. Now, using \eqref{eq:Wpmtransf} we find
\begin{equation}
    \langle 0 | W^{+}(z) \overline{W}^{+}(\overline{\omega})|b_{V,A}(\epsilon)\rangle = 2 \cos(n\epsilon)^2  \langle 0 | W^{+}(z) \overline{W}^{+}(\overline{\omega})|b(0)\rangle \pm 2\sin(n\epsilon)^2 \langle 0 | W^{-}(z) \overline{W}^{-}(\overline{\omega})|b(0)\rangle
\end{equation}
and thus
\begin{equation}
    \omega_{b_{V,A}}(\epsilon) = 2 \cos(n\epsilon)^2  \omega_b(0) \pm 2\sin(n\epsilon)^2\langle 0 | W^{-}(z) \overline{W}^{-}(\overline{\omega})|b(0)\rangle\, ,
\end{equation}
in the last two equations the upper and lower signs refer to the vector and axial phantom symmetry cases. We see that the reflection and transmission coefficients are determined as functions of $\epsilon$ once we known the conformal data of the reference defect. However, due to the presence of an extra $(2,0)$ primary operator besides the stress tensor, it is not enough to have control on the reflection and transmission coefficients of the reference defect, but one needs more detailed information. However, since $W^{\pm}$ are in the same Virasoro representation, it seems unnatural to focus on the 2-point function of only one of them, rather we can define two coefficients $\omega_{b}^{\pm}$ as
\begin{equation}
     \langle 0 | W^{+}(z) \overline{W^{+}}(\overline{\omega})|b\rangle = \frac{\omega_{b}^{\pm}}{(z-\omega)^4}\, ,
\end{equation}
then, along the defect conformal manifold these coefficients change as
\begin{equation}
   \omega_{b}^{\pm}(\epsilon) = 2 \cos(n\epsilon)^2 \omega_{b}^{\pm}(0) + 2 a \sin(n\epsilon)^2 \omega_{b}^{\mp}(0)\, ,
\end{equation}
where $a=1$ ($a=-1$) corresponds to the vector (axial) phantom symmetry. Given those formulas it is simple to derive explicit expression for the reflection and transmission coefficients, the results for a totally transmissive or a factorized reference boundary conditions are reported in \eqref{eq:Rtop} and \eqref{eq:Rfact} respoectively. As an aside curiosity we notice there is a combination of $\omega^{\pm}_b$ that is constant along the conformal manifold
\begin{equation}
    s=\omega_{b}^{-}(\epsilon) + a \omega_b^{+}(\epsilon) = 2 (\omega_{b}^{-}(0) + a \omega_b^{+}(0))\, , 
\end{equation}
thus $s$ is a quantity that characterizes the whole conformal manifold of defects, similarly to the value of the $g$ function. It is simple to understand why $s$ is constant along the conformal manifold. The operators 
\begin{equation}
    \cO^{\pm}(z)= W^{+}(z) \mp i W^{-}(z)
\end{equation}
are eigenstate of the charge $Q$
\begin{equation}
    [Q, \cO^{\pm}(z)] = \pm n \cO^{\pm}(z)\, ,
\end{equation}
and $s$ determines the $2$-point functions of the combinations $\cO^{\pm}(z)\overline{\cO^{\mp}}(\overline{\omega})$ for the vector symmetry and $\cO^{\pm}(z)\overline{\cO^{\pm}}(\overline{\omega})$ for the axial symmetry, both combinations are singlets under the phantom symmetry and the $\epsilon$ independence of $s$ follows.

\end{document}